\pdfoutput=1%tell arXiv to run pdflatex, otherwise hyperref will not work correctly (there is no support for multiline links in .dvi)
\documentclass[twocolumn,english,prb,showpacs,preprintnumbers,amsmath,amssymb]{revtex4-1}
\usepackage[latin9]{inputenc}
\usepackage{esint}
\usepackage{braket}
\usepackage{babel}
\usepackage{tikz}\usetikzlibrary{plotmarks}
\usepackage[colorlinks=true, urlcolor=blue, linkcolor=black, citecolor=black, filecolor=black, menucolor=black, pdfpagelayout=TwoColumnRight, pdfstartview=FitH, plainpages=false, pdfpagelabels,breaklinks=true]{hyperref}
\usepackage{enumitem}

\newcommand{\oncite}[1]{Ref. \onlinecite{#1} }
\newcommand{\oncitep}[1]{Ref. \onlinecite{#1}. }
\newcommand{\oncitec}[1]{Ref. \onlinecite{#1}, }

\begin{document}
\title{ Effective Field Theories for Superconductors in the Subgap Regime}

\author{T.H. Hansson}
\author{T. Kvorning}
\affiliation{Department of Physics, Stockholm University, AlbaNova University
Center, SE-106 91 Stockholm, Sweden}

\author{V.P. Nair}
\affiliation{Physics Department, City College of the CUNY, New York, NY 10031}

\begin{abstract}
We construct  effective field theories for superconductors, that are powerful enough to describe
low lying sub gap fermion modes localized to vortex cores, and at the same time resemble 
topological field theories in that there are no bulk degrees of freedom. 
This is achieved by  a kinetic term for fermions that is proportional to the vortex topological charge, and 
thus vanish in the bulk. 
We study the case of a spin-less two-dimensional $p_x + ip_y$ superconductor in some detail, and show that the 
the subgap fermionic spectrum in a single vortex, including the zero mode, has the same features as those obtained from 
microscopic models. We also show, that  in the 
topological scaling limit our theory becomes a {\em bona fide} topological field theory which retains the 
Majorana modes at the vortex cores, and correctly describes the non-Abelian statistics of such vortices.  
\end{abstract}
\maketitle
%{\em Introduction.}
\section{Introduction}
In topologically ordered phases of matter,
such as quantum Hall (QH) liquids, superconductors, topological insulators
and spin liquids\cite{fradkin13,*bernevig13}, the excitations in the interior of
the system are separated from the ground state by an energy gap, thus
distinguishing them from ordinary metals or magnets. They however
differ in important ways from trivial gapped phases, such as conventional
band insulators, in having excitations with exotic quantum numbers,
and/or gapless edge modes. 
An important theoretical approach to these phases is based on topological
field theories (TFT), which directly builds in important features
of topologically ordered systems, such as the absence of low energy
bulk excitations and, in two-dimensional systems, the possibility
of fractional braiding statistics.

 Prominent examples are the Chern-Simons
(CS) theories of hierarchical QH liquids\, and the BF theories of
superconductors (for reviews, see \cite{wen} and \cite{hos}) and
topological insulators\cite{chomoore,atma}.

That the topological CS theories for the abelian QH liquids encode
the characteristic gapless bosonic edge excitations has been known
for a long time\cite{wenedge}, but, more surprisingly, purely bosonic
TFTs can also describe fermionic edge states. A prominent example
is the Moore-Read, or pfaffian, QH state\cite{mr}, for which a TFT
description based on a $SU(2)$ gauge theory, was proposed by Fradkin
\emph{ et al.}\cite{fradnayak}. 

An important property of the MR state, which it shares with the spinless 2d
  $p_{x}+ip_{y}$ superconductor\cite{readgreen} is that 
 the fundamental vortices support zero energy Majorana modes. 
 As a consequence, a set of $2n$ vortices at fixed positions define a 
 Hilbert space of dimension $2^{n-1}$, and braiding the vortices
 corresponds to unitary rotations in this space. This is the basis
 of the non-abelian fractional statistics that has been looked for in 
 experiments\cite{willett} and is proposed to be useful in quantum 
 information applications\cite{qinfo}.
 
The zero modes in the case of  $p_{x}+ip_{y}$
paired superconductor, is a special example of the vortex subgap modes that occur also 
for s and d-wave pairing. What makes the p-wave case particularly
interesting is that the zero  modes are topologically protected. 
In this context, it is a challenge to formulate effective theories that
describes the physics at energies below the superconducting gap. 
Such theories not only should encode the topological information about
quasiparticles and vortices, but also describe the dynamics of the fermionic  
subgap modes at vortex cores and at edges. The purpose of this letter
is to propose such a theory, and to treat the case of p-wave pairing
of spinless fermions in two dimensions in sufficient detail to 
demonstrate the power of our approach.  

%\textit{The $\psi$BF theory.} 
\section{The $\psi$BF theory}
Our starting point is the topological description of superconductors in terms 
of BF gauge theory which is reviewed in \oncite{hos}. In this theory, the quasiparticle 
current $j_q$ couples to a gauge field $a$ and the vortex current $j_v$ to a 
gauge field $b$. 
The  BF Lagrangian which describes  the topological properties of   superconductors is, 
in the language of differential forms, ${\cal L}_{ BF}=  \frac{1}{\pi} da\, b  -j_{\rm q} a - j_{\rm v} b$. 
In 3d $b_{\mu\nu}$ is an antisymmetric tensor 
field that couples to the world sheet of the propagating vortex string. In  the 2d case, which we will 
concentrate on in the following,  the vortices are point like, and $b_\mu$ is an 
ordinary gauge field.  In standard vector notation we have,
\begin{align}
{\cal L}_{BF}= & \frac{1}{\pi}\epsilon^{\mu\nu\rho}\partial_{\mu}a_{\nu}b_{\rho} -j_{{\rm q}}^{\mu}a_{\mu}-j_{{\rm v}}^{\mu}b_{\mu}\label{bflag}
\end{align}
In addition to the two local  gauge symmetries,  this TFT is also invariant under  parity ($P$) and 
time reversal ($T$).

It is known that by supplementing a TFT with non-topological terms,
scales are introduced and more of the low energy physics can be described.
Adding Maxwell terms to the topological BF theory, introduces both 
a London length, and thus a size for the vortices, and a plasma frequency\cite{hos}.
To describe the subgap fermionic states, we need at least the 
London length, so we shall 
supplement the lagrangian \eqref{bflag} with the  Maxwell terms, 

\begin{align}
	\label{max}
	\mathcal{L}_{M}&=  \frac {\alpha_1} {2\pi} (\vec {E}^{a})^{2}-\frac{\alpha_{2}}{2\pi}(B^{a})^2 +\frac{\beta_{1}}{2\pi}(\vec{E}^{b})^{2}-\frac{\beta_{2}}{2\pi}(B^{b})^{2}
\end{align}
where $B^{b}=\epsilon^{ij}\partial_{i}b_{j}$ \emph{etc.,} and where the
parameters $\alpha_{i}$ and $\beta_{i}$ are related to the London
penetration length $\lambda_{L}$, the Debye screening length $\lambda_{D},$
the plasma frequence $\omega_{p}$ and the vortex energy $\epsilon_{v}$
by $\lambda_{L}=\sqrt{\alpha_{2}\beta_{1}}$, $\lambda_{D}=\sqrt{\alpha_{1}\beta_{2}}$,
$\omega_{p}^{-1}=\sqrt{\alpha_{1}\beta_{1}}$ and $\epsilon_{v}=1/\alpha_{1}$.

Introducing sources, we can solve for the fields in the pure gauge
sector, and for a single, static, pointlike vortex source, $\rho_{v}=\hbar\delta^{2}(\vec{r})$,
we find the solution $B^{a}=\hbar(2\lambda_{L}^{2})^{-1}K_{0}(r/\lambda_{L})$,
$\vec{E}^{b}=-\hbar\alpha_{2}\vec{\nabla}B^{a}$ and $B^{b}=E_{i}^{a}=0$
(in polar coordinates $(r,\theta)$). 

We now present our basic idea. Since the subgap fermion modes are all confined either on the edge of the system, or at the core
of vortices, we want a theory without any bulk fermionic degrees of freedom. We achieve this, not by introducing 
confining potentials, but by having the { \em kinetic energy of the fermions vanish in bulk}. Inspired by \oncite{hks} we make 
the  following ansatz for the fermionic lagrangian, 
\begin{align}
	\mathcal{L}_{\psi}= & \frac{1}{4\pi}\epsilon^{\mu\nu\rho}\partial_{\mu}a_{\nu}\psi^{\dagger} iD_\rho \psi-\tilde{\mathcal{H}} \, ,     \label{mlag}
\end{align}
where $iD_\rho = i\partial_{\rho}+a_{\rho} $. In the case of  $s$-wave  pairing  the fermion field $\psi$ must have
two spin components, while for a spin polarized $p$-wave phase one component suffices. 
For a static vortex, the kinetic term in \eqref{mlag} is $\sim B^a \psi^\dagger \partial_0 \psi$,
which vanishes exponentially outside the vortices. 

The full Hamiltonian is ${\cal H} = - \epsilon_{ij}E^i_a \psi ^\dagger 
i D^j \psi +  \tilde{\mathcal{H}}$, where the first term vanishes for a static vortex. 
$\tilde{\mathcal{H}}$ is to be constructed by a derivative expansion consistent with the 
symmetries of the superconductor in question. 
Also note that there is a natural generalization to 3d by coupling the density 
$\psi^\dagger \gamma_\mu( i\partial_\nu + a_\nu) \psi$ to the topological current $j_v^{\mu\nu} = \epsilon^{\mu\nu\sigma\rho}b_{\sigma\rho}$. 

Combining the pieces \eqref{bflag}, \eqref{max} and \eqref{mlag}  we get 
\begin{align}
{\cal L}_{\psi BF}=\mathcal{L}_{BF}  + \mathcal{L}_{M} +\mathcal{L}_{\psi}    \label{fullag}  
\end{align}
which we shall refer to as the $\psi BF$ Lagrangian. If nodal quasiparticles are present, 
as in the case of \emph{e.g.} $d_{x^2-y^2}$ pairing, extra terms must be added\cite{hermannsmaster}.

To quantize the fermions in \eqref{mlag}, we shall treat $B^{a}$
as a classical background field, to get the commutation relations 
\begin{align}
\left\{ \psi^{\dagger}\left(\vec{r},t\right),\psi\left(\vec{r}^{\prime},t\right)\right\} = & (4\pi/B_{a})\delta^{2}\left(\vec{r}-\vec{r}^{\prime}\right)\label{comrel}
\end{align}
\emph{etc..} Note that $\psi$ is dimensionless, and charged with respect
to the $a$ gauge field.

%{\em The Hamiltonian.} 
\section{The Hamiltonian}
We now proceed to construct $\tilde{\mathcal{H}}$,
so to get a realistic spectrum of subgap modes. In the spirit of effective
field theory, we make a derivative expansion compatible with the symmetries
of the underlying microscopic physics. For a static configuration,
there are two possible terms with no derivatives on the fermion field,
namely $\Lambda\psi^{\dagger}\psi$ and $\mu B^{a}\psi^{\dagger}\psi$.
The first term is crucial for localizing the subgap states at the
vortices, while the second makes no qualitative change and will be
neglected. The quasiparticle states in a superconductor are not charge
eigenstates, and to incorporate this we need a pairing interaction,
which in our case should be of the p-wave type. Since our fermions
are spinless, the lowest derivative pairing interaction possible which
involves only the fields $a$, $b$ and $\psi$, and is invariant
under rotations and gauge transformations, is $\sim\xi E_{z}^{b}\psi\partial_{\bar{z}\,}\psi+h.c.$,
where $z=x+iy$, $E_{z}^{b}=E_{x}^{b}-iE_{y}^{b}$ and $\xi$ is the
the phase operator, introduced by Dirac\cite{dirac}, which can be
used to form a gauge invariant, but non-local, order parameter for
a superconductor, and which transforms as $\xi\rightarrow e^{-2i\zeta}\xi$ under the gauge
transformation $a\rightarrow a+d\zeta$. For details on how to construct $\xi$, see appendix C. In summary, we shall use
\begin{align}
\tilde{\mathcal{H}}= & \Lambda\psi^{\dagger}\psi-\frac{\delta}{4\pi}\xi\left(\vec{r}\right)E_{z}^{b}\psi\partial_{\bar{z}}\psi+h.c.\label{mham}
\end{align}
where $\Lambda$ is an energy density, and $\delta/8\pi$ a dimensionless
coupling parameter. Without loss of generality we can take $\Lambda>0$
and $\delta>0$. Note that the presence of $E^{b}$ in the pairing
term is natural since the current is $\sim\epsilon^{ij}E_{j}^{b}$

\section{The Spectrum}
%{\em The Spectrum}
It is convenient to write the full hamiltonian
in the BdG form, $\mathcal{H}=\frac{1}{2}\Psi^{\dagger}h\Psi$ with
$\Psi^{\dagger}=(\psi^{\dagger},\psi)$, and 
\begin{align}
h= & \frac{1}{2\pi}\begin{pmatrix}h_{0} & \frac{1}{2}\left\{ \partial_{\bar{z}},\Delta\right\} \\
-\frac{1}{2}\left\{ \partial_{z},\Delta^{*}\right\}  & -h_{0}^{*}
\end{pmatrix}\label{oneham}
\end{align}
where $\Delta=\frac{1}{2}\delta\xi E_{z}^{b}$ and, 
\begin{align}
h_{0}= & -\epsilon_{ij}E^i_a i D^j-\Lambda-a_{0}B_{a}\,.\label{elements}
\end{align}
To diagonalize $\mathcal{H}$, we expand the field operators as, $\psi\left(\vec{r},t\right)=\sum_{n}a_{n}(t)u_{n}\left(\vec{r}\right)+a_{n}^{\dagger}(t)v_{n}^{*}\left(\vec{r}\right)$,
and introduce the eigenspinors $\phi\left(\vec{r}\right)=\left(u\left(\vec{r}\right),v\left(\vec{r}\right)\right)^{T}$.
Next we solve the single particle equation for the spinor $\phi$
in the background of widely separated vortices. We begin by removing
the phase of the off-diagonal terms with a gauge transformation. As
can be seen in the supplementary material,  $\xi$ is, in Coulomb gauge,
up to a constant phase, equal to $e^{im\theta}$ for a single vortex
with strength $m$ situated at the origin. So we make the transformation
\begin{align}
\psi\rightarrow & e^{i\frac{1}{2}(m+1)\theta}\,\psi\,,\label{singtrans}
\end{align}
which in general changes the boundary condition on $\psi$. We shall
seek a solution close to the origin, and impose periodic or anti-periodic
boundary conditions on the polar angle, depending on whether the vortex
charge, $m$ at the origin is even or odd\cite{stone,tewari}. For
a static configuration, the other vortices can be neglected, but for
a periodic adiabatic evolution where the vortices encircle each other
the wave function will pick up signs.

The factor $1/B^{a}$ in the commutation relation amounts to having
a modified scalar product for the spinor $\phi$, and this, together
with the singularities in $B^{a}$ and $E_{r}^{b}$ at $r=0$, makes
the eigenvalue problem somewhat subtle. In the supplementary material
it is shown, that for the single particle hamiltonian to be self adjoint,
either $v$ must vanish at $r=0$, or there must exist some real constant $s$ such that
\begin{align}
\lim_{r\rightarrow0}u\left(\vec{r}\right)= & s\lim_{r\rightarrow0}v\left(\vec{r}\right)\,.\label{bcon}
\end{align}
Using the particle hole symmetry of the BdG equations, it follows,
without loss of generality, that a zero energy mode can be written
as $\phi=(\chi,\chi^{*})^{T}$. A direct calculation gives, 
\begin{align}
\chi_{\pm}= & \frac{Ne^{i\left(\frac{\pi}{4}\pm\frac{\pi}{4}\right)}}{\sqrt{\left|E_{r}^{b}\right|r}}\exp\left(\pm\int^{r}dr\,\frac{4\Lambda}{\delta\left|E_{r}^{b}\right|}\right)\,,\label{zerom}
\end{align}
where $N$ is determined by $\int d^{2}r\, B^{a}(r)|\chi(r)|^{2}=2\pi$
which forces $s=\pm 1$ in \eqref{bcon},  depending on the sign of 
$\Lambda$  (which thus has 
no physical meaning). The solution \eqref{zerom} is
an s-wave which exist only for periodic boundary conditions on the
polar angle, \emph{i.e.,} only for odd vortices. {\em We have thus verified
the presence of a zero mode in the correct vortex sector, and at the
same time identified the correct boundary conditions \eqref{bcon}.}

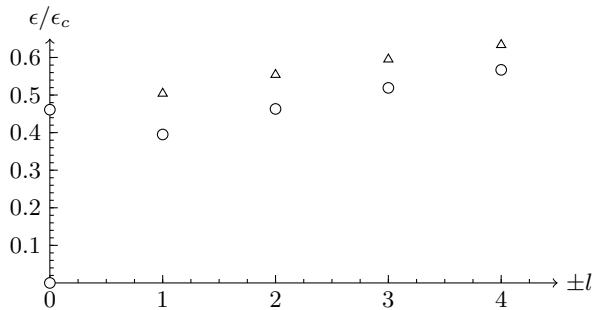
\begin{figure}
\begin{tikzpicture}[only marks, x=1.5cm,y=5cm]   \def\xmin{0}   \def\xmax{4.5}   \def\ymin{0}   \def\ymax{0.65}
  % axes   
\draw[->] (\xmin,\ymin) -- (\xmax,\ymin) node[right] {$\pm l$};   \draw[->] (\xmin,\ymin) -- (\xmin,\ymax) node[above] {$\epsilon/\epsilon_c$};
   %ticks   
\foreach \x in {0,...,4}      	\draw (\x,3pt) -- (\x,0) 		node[anchor=north] {\x};    \foreach \x in {0,0.25,0.5,...,4.25}      	\draw (\x,1.5pt) -- (\x,0);     \foreach \y in {0.1,0.2,0.3,0.4,0.5,0.6}      	\draw (3pt,\y) -- (0,\y)       		node[anchor=east] {\y};      \foreach \y in {0.02,0.04,...,0.62}      	\draw (1.5pt,\y) -- (0,\y); 
  % plot the data  
\draw plot[mark=*, mark options={fill=white}] file {plotlp.data};   \draw plot[mark=triangle*, mark options={fill=white}] file {plotln.data};
\end{tikzpicture}

\caption{The subgap spectrum for $\lambda_{L}=4\lambda_{S}.$ The rings denote
values for positive $l$ and the triangles denote negative. }
\end{figure}

To get the full spectrum of the subgap states, we first neglect the
far away vortices. In this approximation the Hamiltonian is rotation
invariant and we can separate the angular dependence as $\phi\left(\vec{r}\right)=\sum_{l}e^{i\theta l}\phi_{l}\left(r\right)$,
and then determine the radial wave functions and the energy eigenvalues
numerically. Since the kinetic term for the $\psi$ field only has
support on the vortex, which effectively acts as a confining box,
we expect that the spectrum contains an infinite number of bound states
localized at the scale $\lambda_{L}$, but no continuum states. All
our numerical results support this conjecture, and we shall henceforth
assume it to be true; we have not tried to find an analytic proof.
While we do not yet have an analytic proof of this conjecture, all our numerical results
support it and so, henceforth, we shall assume it to be true.

In a type II superconductor the low lying subgap states are however
localized on a smaller scale, $\lambda_{S}$, and for suitable parameters
our model has this feature. 
Indeed, if we introduce the ``confinement'' energy scale $\epsilon_{c}$
by $\Lambda_{h}=\epsilon_{c}/\lambda_{L}^{2}$, an asymptotic analysis
gives $\lambda_{S}=\sqrt{|m|\delta r}\lambda_{L}$ where $r=(\hbar\omega_{p})^{2}/(\epsilon_{v}\epsilon_{c})$.
Thus we can have $\lambda_{S}\ll\lambda_{L}$, by taking a small coupling
parameter $\delta$, and/or making the confinement scale $\epsilon_{c}$
large. In this parameter range we have established numerically that
our spectrum shares important qualitative features with the spectra
obtained by self-consistent solutions of the full microscopic BdG
equations\cite{mizu} (see the figure):
\begin{enumerate}[noitemsep,topsep=-2pt,parsep=2pt,partopsep=0pt,leftmargin=12pt]
	\item The purely angular excitations have a spectrum $E_{0,l}\approx\Delta_{l}+\alpha l$,
with $\alpha\ll\Delta_{l}$. 
	 \item The gap $\Delta_{r}$ for the radial excitations is larger than $\Delta_{l}$. 
	\item The energy scale is inversely proportional to the vortex strength.
\end{enumerate}

The presence of infinitely many bound states is an artifact of our
model, and only the low lying states should be considered as physical.
It is an interesting possibility that adding higher derivative terms
could completely remove the high lying states and leaving a finite
Hilbert space. We leave this as an open problem.

%\noindent{\em The topological scaling limit.}
\section{The topological scaling limit} 
So far, we have shown that the $\psi$BF theory has
all the expected subgap features.
We now show how the theory \eqref{fullag} reduces to a truly 
topological field theory in a proper scaling limit. 
For this, consider  a collection of $N$ identical vortices of unit strength.
The topological scaling limit is defined by taking
both the physical length scale, $\lambda$, and time scale, $\hbar/E$,
to zero at fixed coupling parameters\cite{frohker}. We can think
of $\lambda$ as \emph{e.g.} the minimal distance between the vortices, and
$E$ as a cutoff energy below which our theory is to be valid. 
We define two Majorana fields by, 
\begin{align}
\gamma(\vec{r},t)= & \frac{1}{2}(\psi+\psi^{\dagger}) & \tilde{\gamma}(\vec{r},t)= & \frac{1}{2i}(\psi-\psi^{\dagger})\,,\label{majfield}
\end{align}
and substitute in \eqref{fullag} to get (setting $j_{q}=0$), 
\begin{multline}
\mathcal{L}_{\psi BF}=\frac{1}{\pi}ad(b+\frac{1}{4}\gamma id\gamma+\frac{1}{4}\tilde{\gamma}id\tilde{\gamma})-j_{v}b\\
+\frac{1}{8\pi}ada\psi^{\dagger}\psi-\tilde{\mathcal{H}}\,.\label{eq:subform}
\end{multline}
Since $a_{0}=\vec{E}_{a}=0$, the first term in the second line vanishes.
Also $\vec{E}_{a}=0$ means that $\tilde{{\cal H}}$ is the full hamiltonian,
so the last term in this line is $\sim\sum_{E_{n}<E}E_{n}a_{n}^{\dagger}a_{n}$
which vanishes for fixed $E$ since the subgap (just as all energy scales)
diverges. 
Finally, we make the shift $b\rightarrow b+\frac{1}{4}\tilde{\gamma}id\tilde{\gamma}$
to eliminate the term $ad\tilde{\gamma}id\tilde{\gamma}$ in favor
of $j_{v}\tilde{\gamma}id\tilde{\gamma}$. Using the boundary condition
\eqref{bcon} it is easy to show that $\tilde{\gamma}id\tilde{\gamma}(\vec{0})=0$
so this term vanishes for a point vortex. This concludes the demonstration
that the topological theory,
\begin{align}
\mathcal{L}_{\gamma BF}= & \frac{1}{\pi}\epsilon^{\mu\nu\rho}\partial_{\mu}a_{\nu}\left(b_{\rho}+\frac{1}{4}\gamma\, i\partial_{\rho}\gamma\right)-j_{{\rm q}}^{\mu}a_{\mu}-j_{{\rm v}}^{\mu}b_{\mu} \, ,   \label{tlag}
\end{align}
proposed in \oncitec{hks} is retained in the scaling limit.

%{\em Nonabelian statistics.}
\section{Nonabelian statistics} 
The nonabelian (NA) statistics in
the Moore-Read QH state was originally understood in terms of the
monodromies in the Ising CFT\cite{NW}, assuming that there are no
remaining Berry phases when the wave functions are represented by
conformal blocks. Proofs for this assertion were given in later papers\cite{read/gurbond}.
In the case of the p-wave superconductor, Ivanov\cite{ivanov} derived
the NA statistics using the BdG formulation of Read and Green\cite{readgreen}.
Also here it is important that, in a suitably chosen gauge, there
are no Berry phases, so that the braiding phases of the vortices come
entirely from the coupling to the gauge field. Although quite reasonable,
this is not easy to show, and it was taken for granted by Ivanov.
In a later paper\cite{sternetal} Stern \emph{ et al.} addressed this
question, and gave plausible arguments for the absence of Berry phases
by a more detailed analysis of the vortex cores, using certain mild
assumptions about the continuous part of the spectrum.
In the $\psi$BF theory there can be no Berry phases, since the fermionic
wave functions only have support on the widely separated vortices.
Thus, \emph{ mutatis mutandis}, Ivanov$^{\prime}$s proof of NA statistics
carries over to the $\psi$BF theory, using no extra assumptions. 

We now outline a version of the proof that directly yields the Hilbert space for $2N$ vortices;
details will be given separately\cite{inprep}.
First note that for widely separated vortices, moving along the world lines   $\vec{r}_{a}(t) $, the Majorana field  in  \eqref{majfield} 
takes the form, $\gamma(\vec{r},t)=\sum_{a=1}^{2N}\chi \left(\vec{r}-\vec{r}_{a}(t) \right)\gamma_{a}(t)$
where, in an obvious notation, $\gamma_{a}(t)=a_{0,a}(t)+a_{0,a}^{\dagger}(t)$.
Substituting this in the $\psi$BF Lagrangian, taking the topological scaling limit as above, and using  the normalization
of $\chi$, we retain the following quantum mechanical Lagrangian   
\begin{align}
L_{{\rm M}}= & \frac{m}{4}\sum_{a=1}^{2N}\gamma_{a}(t)i\partial_{t}\gamma_{a}(t)\,,\label{partlag}
\end{align}
where we used the notation, $\gamma_a(t)\equiv\gamma(t,x_{a}^{\mu}(t))$. From \eqref{partlag} follows the 
commutation relations $\{\hat{\gamma}_{a}(t),\hat{\gamma}_{b}(t)\}=2\delta_{ab}$.
Thus, for an adiabatic motion of vortices in the $\psi$BF theory,
by taking the scaling limit, we get a $2N$ dimensional Clifford algebra
at each instant of time. This algebra has a unique irreducible representation
up to a similarity transformation $S$, \emph{i.e.,} $\gamma_{a}(t)=S^{-1}\gamma_{a}S=g_{ab}\gamma_{b}(0)$,
where $g_{ab}\in SO(2N)$. It follows from the connection to the $\psi$BF
model that the operator $S(t)$ is unique and well defined. We can
now express the quantum mechanical Lagrangain \eqref{partlag} in
terms of $g_{ab}$ as 
\begin{align}
L_{M}=-\frac{i}{4}(g^{-1}\dot{g})_{ab}\gamma_{b}(0)\gamma_{a}(0)=-\frac{i}{4}\mathrm{T}r(g^{-1}\dot{g}\, w_{i}^{T}q_{i}) \label{gemact}
\end{align}
where the final form is obtained by bringing the matrix $\gamma_{b}(0)\gamma_{a}(0)$
to canonical form; the weight vector $w_{i}$, which is formed from
the generators of the Cartan subalgebra of $SO(2N)$, depends on the
initial state. Quantizing \eqref{gemact}, using standard methods
based on \cite{nair06,*jedrzej80,*woodhouse97}, the resulting Hilbert space is a representation
space for the spinor representation of $SO(2N)$, which is known to
describe Ising type NA anyons\cite{NW}.

\section{ Final remarks} 
%{\em Final remarks}
 In this letter we proposed  the $\psi$BF theory 
defined by \eqref{fullag} as the proper effective theory for superconductors 
in the energy range below the superconducting gap. 
The distinguishing feature of our theory is that the kinetic term for the fermions has
support only where the vorticity differs form zero, and as consequence, will be confined
to vortex cores. We treated the 2d spinless case
with $p_x + ip_y$ pairing in dome detail, but also indicated how to generalize to 3d and to
other pairing channels.  We are at present working on this. There are also
some aspects of  the 2d case treated here that
remain to be investigated {\em viz,}  the edge excitations, and the ground
state degeneracies on higher genus surfaces. 
Finally, it is a challenge to
derive the $\psi$BF teories from a truly microscopic theory, and also to find
a connection to the Chern-Simons formalism in \oncitep{fradnayak}

\begin{acknowledgments}
We thank M. Sato, A. Stern and X.-L. Qi for interesting discussions and E. Fradkin, A. Karlhede and S. Ryu for helpful
discussions and comments on the manuscript. THH is supported by the
Swedish Research Council, and VPN's work was supported by the U.S.\ National
Science Foundation grant PHY-1213380 and by a PSC-CUNY award.
\end{acknowledgments}
\bibliography{pwave}
\bibliographystyle{apsrev4-1}

\newpage
\clearpage
\newpage
\appendix
\section{The single particle Hamiltonian}

The Lagrangian $\mathcal{L}_{\psi BF}$ gives the canonical equal
time commutation relations 
\begin{align}
\left\{ \psi^{\dagger}\left(\vec{r},t\right),\psi\left(\vec{r}^{\prime},t\right)\right\} = & \frac{4\pi}{B_{a}}\delta^{2}\left(\vec{r}-\vec{r}^{\prime}\right)
\label{eq:comrel}
\end{align}
\emph{etc.,} and the Hamiltonian 
\begin{align*}
H= & \frac{1}{4\pi}\int d^{2}x\begin{pmatrix}\psi^{\dagger} & \psi\end{pmatrix}\begin{pmatrix}H_{0} & -\Delta\partial_{\bar{z}}\\
\Delta^{*}\partial_{z} & -H_{0}^{*}
\end{pmatrix}\begin{pmatrix}\psi\\
\psi^{\dagger}
\end{pmatrix}\,.
\end{align*}
Expanding the fields: 
\begin{align*}
\psi\left(\vec{r},t\right)= & \sum_{n}a_{n}(t)u_{n}\left(\vec{r}\right)+a_{n}^{\dagger}(t)v_{n}^{*}\left(\vec{r}\right)\,,
\end{align*}
and demanding that the mode operators $a_{n}$ shall satisfy $\{a_{m},a_{n}^{\dagger}\}=\delta_{mn}$,
the commutation relations \eqref{eq:comrel} imply the inner product
\begin{align*}
\Braket{\psi|\phi}\equiv & \frac{1}{4\pi}\int d^{2}x\, B_{a}\left(u^{*}U+v^{*}V\right)
\end{align*}
between two single particle states $\psi=\begin{pmatrix}u & v\end{pmatrix}^{T}$
and $\phi=\begin{pmatrix}U & V\end{pmatrix}^{T}$. A straightforward
calculation shows that in order for $a_{n}$ to create a state with
energy $E_{n}$,  to satisfy, $\left[H,a_{n}^{\dagger}\right]=E_{n}a_{n}^{\dagger}$,
there must exist a function $f(\vec{r})$ such that 
\begin{align}
\begin{pmatrix}H_{0} & -\Delta\partial_{\bar{z}}+f(\vec{r})\\
\Delta^{*}\partial_{z}-f(\vec{r}) & -H_{0}^{*}
\end{pmatrix}\begin{pmatrix}u_{n}\\
v_{n}
\end{pmatrix}= & E_{n}B_{a}\begin{pmatrix}u_{n}\\
v_{n}
\end{pmatrix}\,.
\label{modbdg}
\end{align}
This is very similar to the usual single particle BdG equation, but
note the presence of the factor $B_{a}$ multiplying the energy $E_{n}$.
Since $B_{a}$ vanish exponentially away from the vortices, this factor
will drastically change the spectrum, and eliminate the continuum.

We denote the matrix in \eqref{modbdg} by $h$, and since it has
terms that are singular at the origin, some care is needed to define
it properly. 
The problem is reminiscent of giving a proper definition of free anyons,
which mathematically amounts to choosing a particular self-adjoint
extension of the free particle Hamiltonian. In the present case there
turn out to be three possible self-adjoint extensions of the operator
$h$, but (for fixed sign of $\Lambda$) only one of them will support
a zero mode. 
For $h$ to be self-adjoint, it must be symmetric, with respect to
the inner product, \emph{i.e.,} $\forall\phi,\psi\in D_{h}$ 
\begin{align}
\Braket{\psi|H\phi}= & \Braket{H\psi|\phi}\,,\label{symreq}
\end{align}
which implies that $f(\vec{r})=-\frac{1}{2}\left(\partial_{\bar{z}}\Delta\right)$,
and substituting this in \eqref{modbdg} gives eq. \eqref{oneham}
in the main text.
\section{The boundary condition}
\label{boundary}
Since the operator $h$ has terms that are singular at the origin,
there are non-trivial restrictions on its domain $D_{h}$, \emph{i.e.,} on
the allowed boundary conditions at $r=0$, for it to be self-adjoint. 

The condition \eqref{symreq} must be satisfied as $r\rightarrow0$,
but the domain of $h^{\dagger}$ must also be the same as that of
$h$, \emph{i.e.,} $\not\!\exists\psi\in\mathcal{K}/D_{h}$ such that 
\begin{align*}
\Braket{\psi|H\phi}= & \Braket{\psi^{\prime}|\phi}
\end{align*}
for all $\phi\in D_{h}$ and some $\psi^{\prime}\in\mathcal{K}$ (where
$\mathcal{K}$ is the Hilbert space ${\mathrm{L}^2}[R^{2},B_a]\times{\mathrm{L}^2}[R^{2},B_a]$). 

%This mean \emph{e.g.,} that the obvious choice that the wave functions vanish at the origin is a to strong condition
%which have different self-adjoint extensions.

The domain of the Hamiltonian is spanned by the states, 
\begin{align*}
\ket {\psi_l}  = e^{ilm\theta}
\begin{pmatrix}
	e^{-i\frac12 (m+1)\theta}\alpha_l(r)\\
	e^{i\frac12 (m+1)\theta}\beta_l(r)
\end{pmatrix}
\end{align*}
where for modes with $l\neq0$ $\alpha_l(r)$ and $\beta_l(r)$ must vanish at  $r=0$ for  $H \ket {\psi_l} $ to be normalizable. For the $l=0$ modes there is no such condition.
To derive the proper boundary condition we first exclude a disc or
radius $\bar{r}$ around the origin, and then take the limit $\bar{r}\rightarrow0$.
Chosing $\psi=(e^{-i\frac12 (m+1)}A_0(r),e^{i\frac12 (m+1)}B_0(r))$ and $\phi=(e^{-i\frac12 (m+1)}\alpha_0(r),e^{i\frac12 (m+1)}\beta_0(r))$, we have
\begin{align}
 \Braket{H^\dagger \psi|\phi} &\equiv \Braket{\psi|H\phi} \\
   &=  \Braket{H\psi|\phi} -\int d\theta\,\left[\frac{rE_{b}}{2}\delta\left(A_0^{*}\beta_0-B_0^{*}\alpha_0\right)\right]_{r=\bar{r}} \, ,
\label{eq:Hamconstraint}
\end{align}
where the last identity follows by partial integration.
Since $rE_{b}$ remains finite as $r\rightarrow0$ we must have $\lim_{\bar{r}\rightarrow0}\left(A_0^{*}\beta_0-B_0^{*}\alpha_0\right)_{r=\bar{r}}=0$ for $H$ to be self adjoint. To satisfy this condition, we must restrict $D_{h}$. The general solution is
 \begin{align*}
\exists \, s :	\lim_{r\rightarrow0}\alpha_0\left(r\right)&=s\lim_{r\rightarrow0}\beta_0\left(r\right) \ .
\end{align*}
Since this condition holds also  for the $l \neq 0$ modes (they vanish at the origin),  the boundary condition for a general state $\psi=\left(\alpha\left(\vec{r}\right),\beta\left(\vec{r}\right)\right)$ is
\begin{align*}
	\lim_{r\rightarrow0}\alpha\left(\vec{r}\right)&=s\lim_{r\rightarrow0}\beta\left(\vec{r}\right) \ .
\end{align*}
(Note that the obvious choice  \emph{i.e.,} that the wave functions vanish at the origin is a too strong condition in that demanding the integral in 
\eqref{eq:Hamconstraint} to vanish does not at all restrict the domain of $H^\dagger$, meaning that $H$ is not essentially self-adjoint.)
As shown in the main text, only $s=\pm1$ gives  a Hamiltonian which has a zero mode, and the sign is determined by the sign of the coupling $\Lambda$.

%%%%%%%%%%%%%%%%%%%%%%%
\section{The Dirac phase operator}
\label{diracphase}%
The Dirac phase operator $\xi$ related to $a$ transforms as
\begin{align*}
\xi\rightarrow & e^{2i\Lambda}\xi\,,
\end{align*} 
under a gauge transformation $a\rightarrow a+d\Lambda$. In the static case treated in this article we may restrict ourselves
to a gauge sector where $\vec a$ is time independent and $a_{0}=0$. 
For the case of zero magnetic field we can define $\xi$ as the solution to the differential equation
\begin{align}
	-i\vec\nabla\xi=\vec a \xi \ .
	\label{xidef}
\end{align}
If $\vec a$ is sufficiently regular at spatial infinity we can put the boundary condition $\lim_{r\rightarrow\infty\xi}\xi=1$ and we get
\begin{align}
	\xi(\vec{r})= & \exp\left(2i \int d^{2}r^{\prime}\,\vec{a}(\vec{r}^{\prime})\cdot\vec{\nabla}^{\prime}G(\vec{r}^{\prime},\vec{r})\right)\,,
	 \label{lamsol}
\end{align}
where $G$ satisfies $-\vec{\nabla}^{2}G\left(\vec{r},\vec{r}^{\prime}\right)=\delta^{2}\left(\vec{r}-\vec{r}^{\prime}\right)$, and where we also
performed an integration by parts. This is the usual expression for the Dirac phase factor. 

The formula \eqref{xidef} is however not applicable when the  magnetic field is not identically zero. If the magnetic field only had compact support we could define $\xi$ by \eqref{xidef} in the region where the magnetic field is zero, and then analytically continue to the whole plane. In the case relevant for this article the magnetic field does not have  compact support, but is exponentially localized with a localization length $\lambda_L$ around points $\{\vec r_a\}$. The straight forward forward generalization to this case would then be
\begin{align}
	-i\vec\nabla\xi=\lim_{\lambda_L\rightarrow 0}\vec a_{\lambda_L} \xi \ ,
	\label{xidef2}
\end{align}
but then we have to specify a family of vector potentials $\{\vec a_{\lambda_L}\}$. This family should only be defined by the requirement that no gauge transformation is associated with the change of $\lambda_L$. More precisely, the equation
\begin{align*}
\int d^{2}r^{\prime}\,\vec a_{\lambda_L}(\vec{r}^{\prime})\cdot\vec{\nabla}^{\prime}G(\vec{r}^{\prime},\vec{r})=\int d^{2}r^{\prime}\,\vec a_{\lambda_L^\prime}(\vec{r}^{\prime})\cdot\vec{\nabla}^{\prime}G(\vec{r}^{\prime},\vec{r}) 
\end{align*}
should hold for all $\lambda_L$ and $\lambda_L^\prime$.

Returning to the vortex configuration considered in the text, and using Coulomb gauge ,we have
\begin{align*}
\vec{a}= m\hat{\theta}\left(\frac{1}{r}-\frac{1}{\lambda_{L}}K_{1}\left(\frac{r}{\lambda_{L}}\right)\right)
= m\left[ \vec \nabla \theta -\frac{1}{\lambda_{L}}K_{1}\left(\frac{r}{\lambda_{L}}\right)  \right]
\end{align*}
for a vortex of strength $m$. Since $\lim_{\lambda_L\rightarrow0}\frac{1}{\lambda_{L}}K_{1}\left(\frac{r}{\lambda_{L}}\right)=0$ we can read of the solution to \eqref{xidef2}, and we get $\xi = e^{i(m \theta +\alpha)}$, with $\alpha$ being a real constant which we, without loss of generality, can put to zero.

\section{Numerics}

To get the spectrum of the single particle Hamiltonian we project
it to a finite Hilberts space in which we diagonalize exactly. Because
of the rather unusual inner product (with a measure $\sim B_{a}\sim K_{0}$),
usual basis sets, such as cylindrical waves, will give a generalized
eigenvalue problem with a matrix with a very small determinant, implying
that the overlap integrals have to be calculated with a very high
precision.

To overcome this problem one can either try to find a basis which
from the start is close to orthogonal, w.r.t. the inner product, and
has sizable matrix elements for the matrix $h$, or to fine a basis
where all integrals can be analytically, and thus be evaluated to
a very high precision.

We choose the the second strategy and used the basis functions $\phi_{a}=r^{a}\sqrt{rE^{b}(r)}$
since all relevant matrix elements can be evaluated using the formula, 
\begin{widetext}
\begin{align*}
\int_{0}^{\infty}r^{a}K_{b}(r)K_{c}(r)dr= & \frac{\Gamma\left(\frac{1}{2}(a-b+c+1)\right)\Gamma\left(\frac{1}{2}(a+b+c+1)\right)\Gamma\left(\frac{1}{2}(a-b-c+1)\right)\Gamma\left(\frac{1}{2}(a+b-c+1)\right)}{2^{2-a}\Gamma(a+1)}
\end{align*}

\end{widetext}
To estimate the precision of our result, we increased the maximum
exponent $a_{max}$, from 125 to 250, which gave a relative changes
in the eigenvalues of at most $<10^{-2}$, as claimed in the main
text.
\end{document}